\documentclass[aps,prd,showpacs,preprint,floatfix,nofootinbib]{revtex4}
\usepackage{amsmath}
\usepackage{amstext}
\usepackage{amsthm}
\usepackage{amssymb}
\usepackage{graphicx}

\newcommand{\be}{\begin{equation}}
\newcommand{\ee}{\end{equation}}
\newcommand{\ba}{\begin{eqnarray}}
\newcommand{\ea}{\end{eqnarray}}
\newcommand{\bc}{\begin{center}}
\newcommand{\ec}{\end{center}}

\def\rp{r_{+}}
\def\re{r_{\rm EH}}

\begin{document}

\preprint{MZ-TH/06-04}
\begin{center}

{\LARGE \textsc{\\}}
{\LARGE \textsc{\\}}
{\LARGE \textsc{Spacetime structure of an evaporating black hole in quantum gravity}}

\vspace{1.4cm}
{\large A.~Bonanno}\\

\vspace{0.7cm}
\noindent
\textit{INAF -  Osservatorio Astrofisico di Catania,
Via S.Sofia 78, I-95123 Catania, Italy\\
INFN, Via S. Sofia 64, I-95123 Catania, Italy}\\

\vspace{0.7cm}
{\large M.~Reuter}\\

\vspace{0.7cm}
\noindent
\textit{Institute of Physics, University of Mainz\\
Staudingerweg 7, D-55099 Mainz, Germany}\\

\end{center}
\vspace*{0.6cm}
\date{\today}

\begin{abstract}
The impact of the leading quantum gravity effects on the dynamics of the Hawking evaporation
process of a black hole is investigated. Its spacetime structure is described by a renormalization group
improved Vaidya metric. Its event horizon, apparent horizon, and timelike limit surface are obtained 
taking the scale dependence of Newton's constant into account. The emergence of a quantum
ergosphere is discussed. The final state of the evaporation process is a cold, Planck size remnant.
\end{abstract}

\maketitle

\renewcommand{\theequation}{1.\arabic{equation}}
\setcounter{equation}{0}

\section{introduction}
One of the very remarkable features of black hole radiance  \cite{hawking} is the observation that the
global spacetime structure of a black hole losing mass by the evaporation process is far more complicated 
than that of its static counterpart \cite{ybook,y1,y2}. 
Even a Schwarzschild black hole, when it radiates, does 
not have a single horizon that fully characterizes its structure, and 
one must distinguish at least three important horizon-like loci. 
The future event horizon (EH) is the boundary of the causal past of future null infinity, 
and it represents the locus of outgoing future-directed null geodesic
rays that never manage to reach arbitrarily large distances from the hole.
The apparent horizon (AH) is defined as the outermost marginally trapped surface for the 
outgoing photons. Classically it can be null or spacelike, in presence of quantum radiance it can be timelike 
also,  when regarded as a 3-dimensional surface. The third important locus is the timelike limit surface
(TLS) or ``quasi-static limit"  which is defined as the locus where static observers become lightlike.
The TLS can be null, timelike, or spacelike \cite{ybook}.
For a classical Schwarzschild black hole (which does not radiate), the three surfaces EH, AH, and TLS are all
identical. Upon ``switching on" the Hawking evaporation this degeneracy is partially lifted. According to
the analysis by York \cite{ybook,y1} the AH continues to coincide with the TLS for a spherically 
symmetric emission, but the EH is different from AH=TLS.

In particular, if we approximate the stress-energy tensor near the horizon as a radial
influx of negative energy which balances the outward Hawking flux at infinity,
the event horizon is  located {\it inside} the AH \cite{is}, the portion of spacetime between the two surfaces 
forming the so-called ``quantum ergosphere". 
This name stems from the analogy with the classical  (stationary) Kerr black hole for which 
EH=AH$\not =$TLS. Here the ergosphere is the space between ``the" horizon EH=AH and the TLS,
usually called the ``static limit". In both cases particles and light signals can escape from within the 
ergosphere and reach infinity.

The definition of the EH via the locus of outgoing photons 
that can never reach large distances from the hole has the unfortunate  ``teleological" 
property of requiring knowledge of the entire future history of the 
hole \cite{ybook, poi}. In particular when the black hole radiance is described semiclassically 
(quantized matter in a classical geometry), the Bekenstein-Hawking temperature and the luminosity diverge
for $M\rightarrow 0$, as $T_{\rm BH} \propto 1/M$ and $L\propto 1/M^2$, respectively. As a result, 
this approximation breaks down for very light holes, and in order to determine the final state 
of the evaporation process a much more precise treatment, including backreaction and quantum gravitational
effects, is required.

In York's work \cite{ybook,y1}, which is strictly within the semiclassical approximation, 
the ``teleological" problem is
circumvented by relaxing the definition of the EH in the following way.  Rather than demanding that the 
photons ``never" reach infinity he demands only that they are imprisoned by the event horizon for times
which are very long compared to the dynamical time scale of the hole. Using this working definition
of the EH he is then able to determine its location to first order in the luminosity $L$. In this manner the 
difficult question about the real final state of the evaporation is not touched upon. 

It is the purpose of the present paper to analyze the dynamical evaporation process and the corresponding
spacetime structure of a radiating Schwarzschild black hole. We include the leading quantum gravitational
corrections of the geometry which, as we shall discuss, seem to lead to a termination of
the evaporation process and the formation of a cold, Planck size remnant. 
Our main tool will be the ``renormalization group improvement" of classical solutions, a technique which is  very popular
in conventional field theory.

In fact, recently a lot of work went into the investigation of the nonperturbative renormalization group
(RG) behavior of Quantum Einstein Gravity 
\cite{mr,percadou,oliver1,frank1,oliver2,oliver3,oliver4,oliver5,oliver6,souma,percacciperini,
percaper2,frank2,frankf,litimgrav,max,max2,max3,brproper,resh}
and its possible manifestations \cite{bh1,bh2,cosmo1,cosmo2,elo,esposito,scalfact,bauer,h1,h2,wey70}. 
In particular, in \cite{bh2}, a ``RG-improvement" of the Schwarzschild metric has
been performed and the properties of the corresponding ``quantum black hole" have been explored. The improvement
was based upon the scale dependent (``running") Newton constant $G(k)$ obtained from the exact RG
equation for gravity \cite{mr} describing the scale dependence of the effective average action \cite{avact,ym1,ym2,ym3,ym4}.
Here $k$ denotes the mass scale of the infrared cutoff which is built into the effective average action
$\Gamma_k [g_{\mu\nu}]$ in such a way that it generates the field equations for a metric which has been 
averaged over a spacetime volume of linear dimension $k^{-1}$. The running of $G$ is approximately given by 
\be\label{2.23}
G(k)={G_0\over 1+\omega \; G_0 \; k^2}
\ee
where $G_0$ denotes the laboratory value of Newton's constant, and $\omega$ is a constant. At large distances 
$(k\rightarrow 0)$, $G(k)$ approaches $G_0$, and in the ultraviolet limit $(k\rightarrow \infty)$, 
it decreases as $G(k)\propto 1/k^2$. This is the fixed point behavior responsible for the conjectured 
nonperturbative renormalizability of Quantum Einstein Gravity \cite{mr,oliver1,oliver2,souma}. 

In the RG improvement
scheme of \cite{bh2} the information about the $k$-dependence of $G$ is exploited in the following way.
The starting point is the classical Schwarzschild metric (in Schwarzschild coordinates)
\be\label{due}
ds^2 = -f(r) dt^2 + f(r)^{-1}dr^2 + r^2 d\Omega^2
\ee
with $d\Omega^2 \equiv d\theta^2 +\sin^2 \theta d\phi^2$ and the classical lapse function
$f(r)= 1-2G_0 M/r \equiv f_{\rm class}(r)$. The RG improvement is effected by substituting, 
in $f_{\rm class}(r)$, $G_0$ by the $r$-dependent Newton constant $G(r)\equiv G(k=k(r))$ which 
obtains from $G(k)$ via an appropriate ``cutoff identification" $k=k(r)$. In flat space 
the natural choice would be $k \propto 1/r$. In \cite{bh2} we argued that in the 
Schwarzschild background the correct choice,   
in leading order at least, is $k(r)= \xi/d(r)$ where $\xi$ is a constant of the 
order of unity, and $d(r)\equiv \int_0^r dr' | f_{\rm class}(r')|^{-1/2}$ is the proper distance from a point
with coordinate $r$ to the center of the black hole. (We refer to \cite{bh2} for a detailed physical
justification of this choice.) While the integral defining $d(r)$ can be evaluated exactly, it is sufficient
to use the following approximation which becomes exact for both $r\rightarrow \infty$ and $r \rightarrow 0$: 
\be\label{tre}
d(r)=\left ( {r^3\over r+\gamma \; G_0\; M} \right )^{1\over 2}
\ee
The resulting $G(r)\equiv G(k=\xi/d(r))$ reads
\be\label{quattro}
G(r)={G_0 \; r^3\over r^3 +\tilde{\omega}\;G_0\; [r+\gamma G_0 M]}
\ee
where $\widetilde{\omega}\equiv \omega\xi^2$. In these equations the parameter $\gamma$ has the value
$\gamma =9/2$ if one sets $k=\xi/d(r)$ as above. It turns out, however, that most of the
qualitative properties of the improved metric, in particular all those related to the structure of its horizons,
are fairly insensitive to the precise value of $\gamma$. In particular, $\gamma=0$ (corresponding to $k=\xi/r$) 
and $\gamma =9/2$ where found \cite{bh2} to lead to rather similar results throughout.
For this reason we shall adopt the choice $\gamma=0$ in the present paper. It has the advantage that with this choice
many calculations can be performed analytically which require a numerical treatment otherwise.

The metric of the RG improved Schwarzschild black hole is given by the line element (\ref{due}) with 
\be\label{cinque}
f(r)=1-\frac{2 G(r) M}{r}
\ee
Let us briefly list its essential features\footnote{All formulas quoted refer to $\gamma=0$, but the 
qualitative features are the same for $\gamma =9/2$; see \cite{bh2} for details.}.

a) There exists a critical mass value\footnote{We define the (standard) Planck mass and length 
in terms of the laboratory value $G_0$: $m_{\rm Pl}=\ell_{\rm Pl}^{-1}=1/\sqrt{G_0}$.}
\be\label{sei}
M_{\rm cr}=\sqrt{\widetilde{\omega}/G_0}=\sqrt{\widetilde{\omega}} \;  m_{\rm Pl}
\ee
such that $f(r)$ has two simple zeros at $r_{-}$ and $r_{+}>r_{-}$ if $M>M_{\rm cr}$, one double zero
at $r_{+}=r_{-} = \sqrt{\widetilde{\omega}G_0}$ if $M=M_{\rm cr}$, and no zero at all if $M<M_{\rm cr}$.
For $M>M_{\rm cr}$ the zeros are at 
\be\label{sette}
r_{\pm} = G_0 M \; [1\pm \sqrt{1-\Omega}]
\ee
with the convenient abbreviation 
\be\label{otto}
\Omega \equiv \frac{M_{\rm cr}^2 }{M^2} = \widetilde{\omega} \; \Big ( \frac{m_{\rm Pl}}{M} \Big )^2 
\ee
The spacetime has an outer horizon at $r_{+}$ and in inner (Cauchy) horizon at $r_{-}$ . At 
$M_{\rm cr}$, the black hole is extremal, the two horizons coincide, and
the spacetime is free from any horizon if the mass is sufficiently small,  $M<M_{\rm cr}$.

b) The Bekenstein-Hawking temperature $T_{\rm BH}= \kappa /2\pi$ is given by the surface gravity 
at the outer horizon, $\kappa = {1\over 2}f'(r_{+})$. Explicitly,
\be\label{nove}
T_{\rm BH}(M)  = {1\over 4\pi G_0 M}\;{\sqrt{1-\Omega}\over 1 +\sqrt{1-\Omega}}
={1\over 4\pi G_0 M_{\rm cr}} \; {\sqrt{\Omega(1-\Omega)} \over 1 +\sqrt{1-\Omega}}= 
\frac{M_{\rm cr}}{4\pi\widetilde{\omega}}
{\sqrt{\Omega(1-\Omega)}\over 1+\sqrt{1-\Omega}}
\ee
This temperature vanishes for $M \searrow M_{\rm cr}$, 
{\it i.e.} $\Omega \nearrow 1$, thus motivating the interpretation of the improved Schwarzschild 
metric with $M=M_{\rm cr}$ as describing a ``cold" remnant of the evaporation process.

c) The energy flux from the black hole, its luminosity $L$, can be estimated using Stefan's law. 
It is given by $L=\sigma {\cal A}(M) T_{\rm BH}(M)^4$ where $\sigma$ is a constant and ${\cal A}\equiv
4\pi r_{+}^2$ denotes the area of the outer horizon. With (\ref{sette}) and (\ref{nove})
we obtain
\be\label{dieci}
L(M) = {\sigma \; M_{\rm cr}^2\over (4\pi)^3 \; \widetilde{\omega}^2}\;
{\Omega (1-\Omega)^2\over [1+\sqrt{1-\Omega}]^2}
\ee 
For a single massless field with two degrees of freedom one has $\sigma = \pi^2 /60$.
(We use units such that $\hbar=c=k_B=1$.) 
\renewcommand{\theequation}{2.\arabic{equation}}
\setcounter{equation}{0}
\section{The quantum-corrected Vaidya metric}
Our aim is to find a metric which describes the history of an evaporating Schwarzschild black hole
and its gravitational field. In the small luminosity limit $(L\rightarrow 0)$ this metric is supposed 
to reduce to the static metric of the RG improved Schwarzschild spacetime.

We begin by reexpressing the metric (\ref{due}) with the improved lapse function (\ref{cinque}) in terms
of ingoing Eddington-Finkelstein coordinates $(v,r,\theta,\phi)$. We trade the Schwarzschild time $t$
for the advanced time coordinate 
\be\label{dueuno}
v=t+r^\star, \;\;\;\;\;\;\;\;\;\;\;\; r^\star \equiv \int^r dr' /f(r')
\ee
Here $r^\star$ is a generalization of the familiar ``tortoise" radial coordinate to which it 
reduces if $G(r)=const$. For $G(r) \not = const$ the function $r^\star= r^\star(r)$ is more complicated, but its
explicit form will not be needed here. Eq.(\ref{dueuno}) implies 
$dv= dt+dr/f(r)$, turning (\ref{due}) with (\ref{cinque}) into 
\be\label{duedue}
ds^2=-[1-2G(r)M/r] \; dv^2 + 2 dv dr +r^2d\Omega^2
\ee
Eq.(\ref{duedue}) is exactly the Schwarzschild metric in Eddington-Finkelstein coordinates, with $G_0$
replaced by $G(r)$. It is thus reassuring to see that the two operations, the RG improvement 
$G_0\rightarrow G(r)$ and the change of the coordinate system, can be performed in either order, they 
``commute".

The thermodynamical properties derived in \cite{bh2} and summarized in the previous section refer to the 
metric (\ref{duedue}). In the exterior of the hole the spacetime is static, 
and while we can deduce a temperature and a corresponding luminosity from its periodicity in imaginary time 
(or by computing the surface gravity  at  $r_{+}$ directly) the backreaction of the mass-loss due to the evaporation
is not described by (\ref{duedue}). From the static metric we obtained the mass dependence of the 
luminosity, $L=L(M)$. Using this information we can compute the mass of the hole as seen by 
a distant observer at time $v$, $M(v)$, by solving 
the differential equation
\be\label{duetre}
-{d \over dv} M(v) = L(M(v))
\ee
In our case $L(M)$ is given by Eq.(\ref{dieci}). 
To first order in the luminosity, the metric which incorporates the effect of the decreasing mass is obtained by 
replacing the constant $M$ in (\ref{duedue}) with the $M(v)$ obtained from Eq.(\ref{duetre}):
\be\label{duequattro}
ds^2=-[1-2G(r)M(v)/r] \; dv^2 + 2 dv dr +r^2d\Omega^2
\ee
For $G(r)=const$, Eq.(\ref{duequattro}) is the Vaidya metric which frequently had been used to explore the influence of the Hawking 
radiation on the geometry \cite{bar,his,his2}. It is a solution of Einstein's equation $G_{\mu\nu}=8\pi G_0 T_{\mu\nu}$
where $T_{\mu\nu}$ describes an inward moving null fluid. In this picture the decrease of $M$ is due to the
inflow of negative energy, as it is appropriate if the field whose quanta are radiated off is in the Unruh vacuum \cite{bd}.

The metric (\ref{duequattro}) can be regarded as a RG improved Vaidya metric. It encapsulates two different mechanisms 
whose combined effect can be studied here: the black hole radiance, and the modifications of the spacetime structure due to the 
quantum gravity effects, the running of $G$ in particular. 

It is instructive to ask which energy-momentum tensor $T_{\mu\nu}$ would give rise to the improved
Vaidya metric (\ref{duequattro}) according to the classical equation ${G_{\mu}}^{\nu}=8\pi G_0 {T_{\mu}}^{\nu}$. Computing the 
Einstein tensor of (\ref{duequattro}) one finds that its only non-zero components are
\begin{subequations}\label{2.5}
\ba
&& {T^{v}}_v = {T^r}_r=-\frac{G'(r)M(v)}{8\pi G_0 r^2}\label{2.5a}\\[2mm]
&& {T^r}_v = \frac{G(r) \dot{M}(v)}{8\pi G_0 r^2}\label{2.5b}\\[2mm]
&& {T^{\theta}}_\theta =  {T^\phi}_\phi= -\frac{G''(r) M(v)}{16\pi G_0 r} 
\ea
\end{subequations}
Here the prime (dot) denotes a derivative with respect to $r(v)$. The non-zero components (\ref{2.5}) contain either $r$- or 
$v$-derivatives but no mixed terms. The terms with $r$-derivatives of $G$, also present for $M(v)=const$, describe the
vacuum energy density and pressure of the improved Schwarzschild spacetime in absence of radiation effects. 
They have been discussed in \cite{bh2} already. Allowing for $M(v)\not = const$, the new feature is a nonzero component ${T^r}_v\not = 0$
which, for $\dot {M}< 0$, describes the inflow of negative energy into the black hole.

\begin{figure}
\includegraphics[width=12cm]{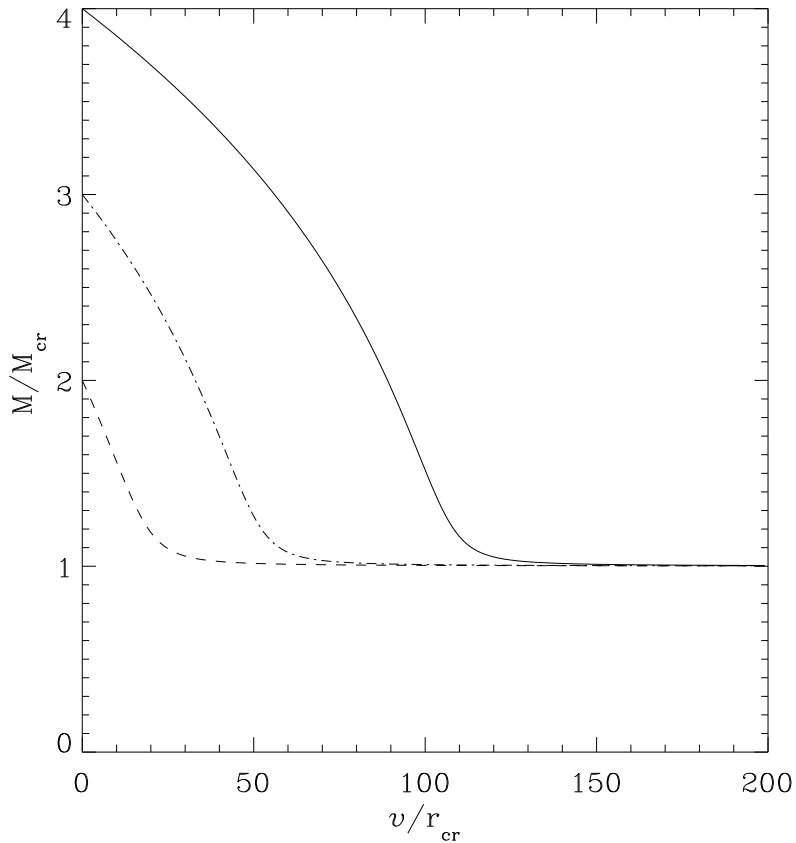}
\caption{\label{fig1}
The ratio $M/M_{\rm cr}$ as a function of $v/r_{\rm cr}$  
for various initial masses. } 
\end{figure}

Taking advantage of the luminosity function $L(M)$, 
Eq.(\ref{dieci}), we can solve the differential equation (\ref{duetre})  
numerically and obtain the mass function $M=M(v)$. (We have 
set $\sigma / (4\pi)^3 \widetilde{\omega} = 1$ in the 
numerical calculations in order to reach the almost complete evaporation  
for $v\approx 200$ in units of $r_{\rm cr}$.) 
The result is shown in Fig.(\ref{fig1}) for various initial masses, in the domain $v>0$. In fact, 
for definiteness we assume that the black hole is formed at $v=0$ by the implosion of a spherical
null shell \cite{his}. Hence $M(v)$ is given by Fig.(\ref{fig1}) together with $M=0$ for $v<0$. We observe that,
for any initial mass, $M(v)$ approaches the critical mass $M_{\rm cr}$ for $v\rightarrow \infty$.
This behavior is the most important manifestation of the quantum gravity effects: according to 
Eq.(\ref{nove}), the temperature $T_{\rm BH}(M)$ goes to zero when $M$ approaches $M_{\rm cr}$ from
above. Hence the luminosity vanishes, too, the evaporation process stops, and $M(v)\approx M_{\rm cr}$ remains
approximately constant at very late times, $v \gg M_{\rm cr}^{-1}$. 

\begin{figure}
\includegraphics[width=12cm]{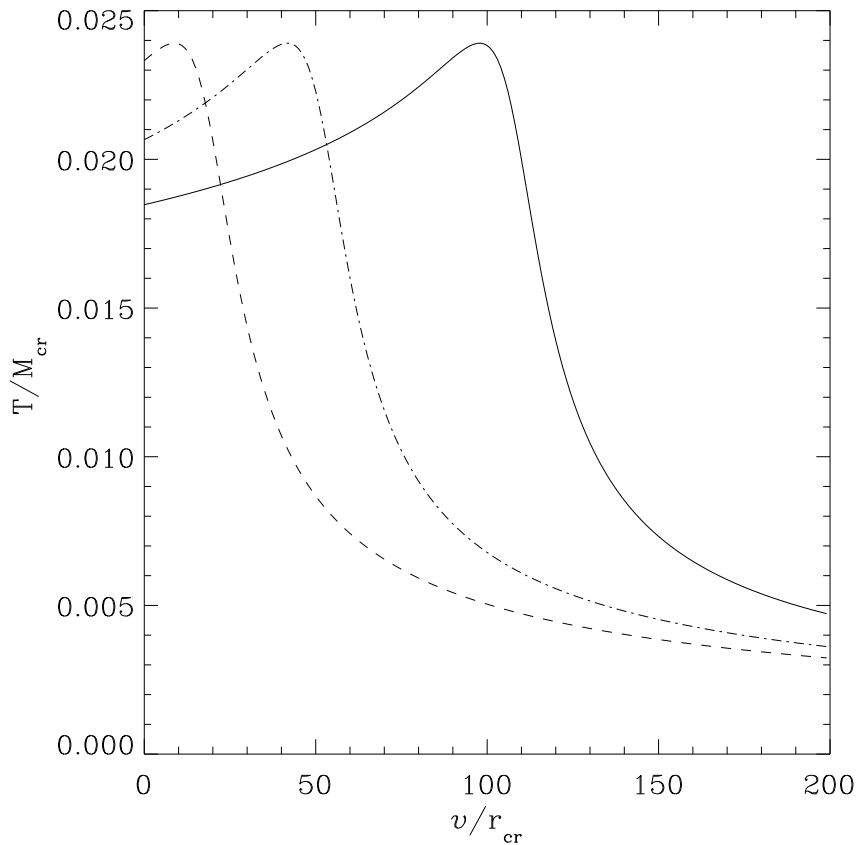}
\caption{\label{fig2}
Time dependence of the Bekenstein-Hawking temperature during the evaporation process for the same 
initial masses as in Fig.(\ref{fig1}).}
\end{figure}

\begin{figure}
\includegraphics[width=12cm]{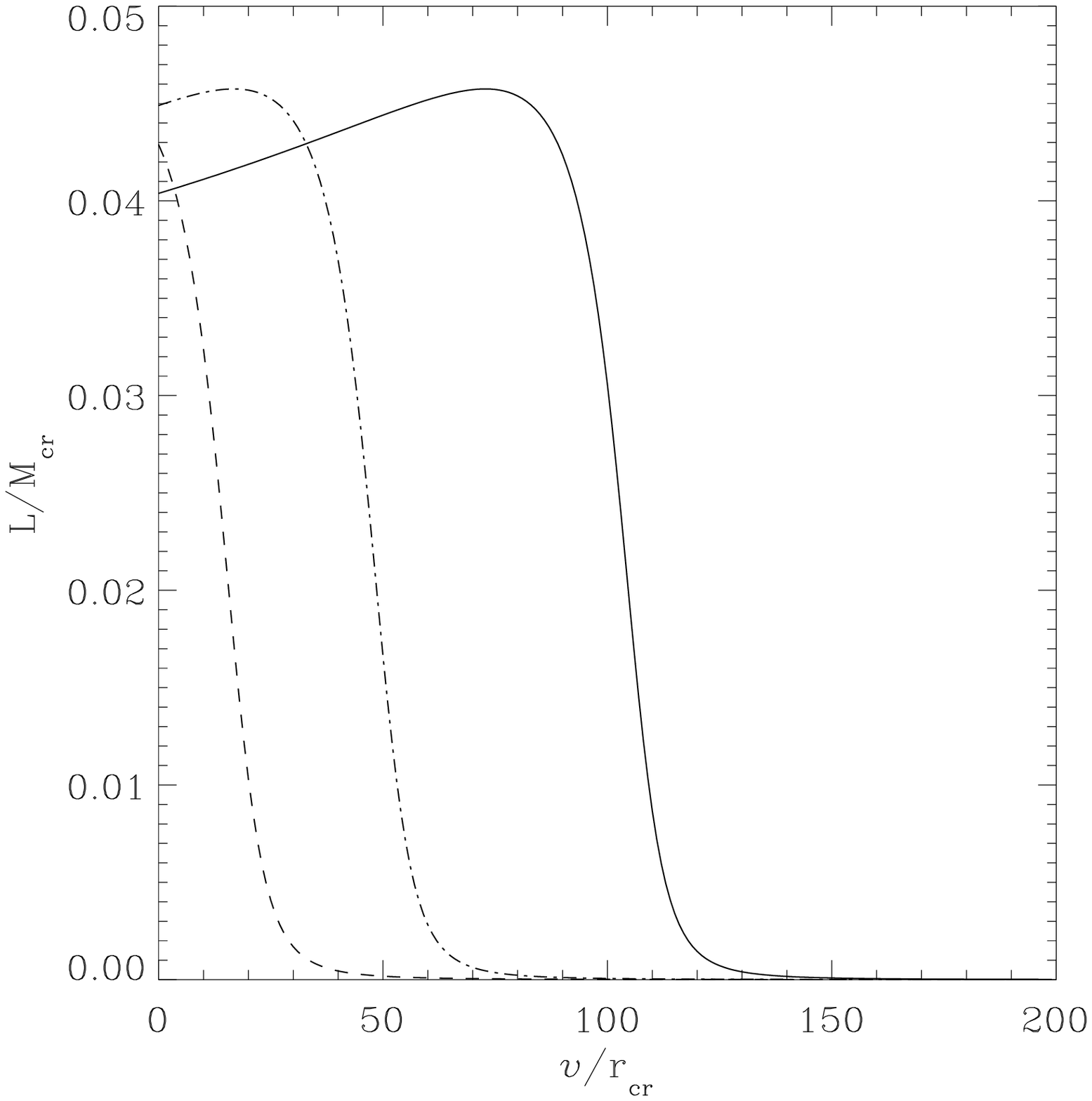}
\caption{\label{fig3} 
The black hole's luminosity as a function of $v/r_{\rm cr}$, for the same initial
masses as in Fig.(\ref{fig1}) and Fig.(\ref{fig2}).}
\end{figure}

In Fig.(\ref{fig2}) and Fig.(\ref{fig3}) we plot the advanced time dependence of the temperature
$T_{\rm BH}(v)\equiv T_{\rm BH}(M(v))$ and the luminosity $L(v)\equiv L(M(v))$, respectively.
They are obtained by inserting the numerical solution of Eq.(\ref{duetre}) into (\ref{nove})
and (\ref{dieci}). 

Both the very early and the very late stages of the evaporation process can be described analytically.
A black hole with $M(v=0)\gg M_{\rm cr}$ 
starts in what we call the ``Hawking regime". It 
is defined by the approximation $\Omega \approx 0$ which is realized
if the hole is very heavy ($M\gg M_{\rm cr}$)
or if $\widetilde{\omega} =0 $ in the semiclassical limit
where the quantum gravity corrections are ``switched off".
In the Hawking regime, (\ref{nove}) and (\ref{dieci}) reproduce the familiar results
\begin{subequations}\label{2.6}
\ba
&& T_{\rm BH} (M)= \frac{1}{8\pi G_0 \; M} \label{2.6a}\\[2mm]
&& L(M) = \frac{B}{G_0^2 \; M^2}, \;\;\;\;\;\;\;\;\;\;\; B\equiv\frac{\sigma}{4(4\pi)^3}  \label{2.6b}
\ea
\end{subequations}
It is easy to solve the differential equation $-\dot{M}=L(M)$ for the luminosity (\ref{2.6b}). With the initial 
condition $M(v=0)=M_0$ the solution reads
\be\label{duesette}
M(v)=\Big [M_0^3-3(B/G^2_0) \;v \Big ]^{1/3}.
\ee
This is the mass function during the early stages of the evaporation process, valid as long as $M(v)$
is well above the critical mass. If one naively extrapolates (\ref{duesette}) to small masses one finds  $M(v_0)=0$, 
implying a final ``explosion" with $T\rightarrow \infty$ and $L\rightarrow \infty$, after a finite time 
$v_0 = G_0^2 M_0^3 / (3 B)$. As a consequence of the quantum gravity effects, this is not what really happens, however.

The final part of the evaporation process, where the cold remnant forms, is in the ``critical regime". It is 
described by those terms in the above expressions which are dominant for $M\searrow M_{\rm cr}$, or $\Omega \nearrow 1$. From (\ref{nove})
and (\ref{dieci}) we obtain in this approximation:
\begin{subequations}
\ba
&& T_{\rm BH}(M) = \frac{1}{4\pi \widetilde{\omega}}\sqrt{M^2-M^2_{\rm cr}} \label{2.8a}\\[2mm]
&& L(M) = \frac{\sigma G_0}{(4\pi\widetilde{\omega})^3}\bigg (M^2-M^2_{\rm cr} \bigg )^2  \label{2.8b}
\ea
\end{subequations}
Solving $-\dot{M} = L(M)$ with (\ref{2.8b}) one finds
\be\label{duenove}
M(v)=M_{\rm cr}+\frac{M_1 -M_{\rm cr}}{1+\alpha (M_1-M_{\rm cr})(v-v_1)}
\ee
Here $\alpha\equiv \sigma/(16\pi^3\widetilde{\omega}^2)$, and $v_1$ is a time, already in the 
critical regime, where $M(v_1)=M_1$ is imposed.
For $v\rightarrow \infty$, the difference $M(v) - M_{\rm cr}$ vanishes proportional to $1/v$, as 
a consequence of which $T_{\rm BH}(v)\rightarrow 0$ and $L(v)\rightarrow 0$.
 
We mentioned already that the RG improved Vaidya metric (\ref{duequattro}) can be a correct description
only to first order in $L$. In fact, deriving the surface gravity and luminosity from (\ref{duequattro}) the results differ from
those for the improved Schwarzschild metric by terms due to the $v$-dependence of $M$. In our approximation those
terms are neglected as they would contain additional factors of $\dot{M} = -L$. 
\renewcommand{\theequation}{3.\arabic{equation}}
\setcounter{equation}{0}
\section{apparent horizon and timelike limit surface}
Next we turn to the various horizon-like loci of the improved Vaidya metric. Regarded as $3$-surfaces, all of them are
histories of spherical $2$-surfaces.  

The apparent horizon is a marginally trapped surface. We determine it from the condition that one of the congruences of 
radial null geodesics, in affine parametrization, has vanishing expansion scalar there, $\Theta=0$. The improved Vaidya
metric (\ref{duequattro}) has the structure
\be\label{treuno}
ds^2=[-f(r,v)dv+2dr]dv+r^2d\Omega^2
\ee
Along outgoing radial null geodesics we have $f dv = 2dr $. Hence, parametrizing them as $r=r(v)$, they obey the differential
equation $\dot{r}(v)=f \big (r(v),v \big)/2$ where the dot denotes a derivative with respect to $v$. We can rewrite this equation in the
autonomous form
\be\label{tredue}
\frac{d }{d\lambda} x^\mu(\lambda)= u^\mu (x(\lambda))
\ee
with the null vector field
\be\label{tretre}
u^{\mu}\equiv (u^v,\; u^r, \; u^\theta, \; u^\phi)=(1,\;{1\over 2} f, \;0, \; 0 )
\ee
A short calculation reveals that the geodesic equation holds in the 
form $u^\nu D_\nu u^\mu= {\cal K}u^\mu$ with a nonzero function 
\be\label{trequattro}
 {\cal K} = \frac{1}{2} \partial_r f
\ee
Hence the parameter $\lambda$ in (\ref{tredue}) is not an affine one. In order for the standard discussion \cite{poi,he}
to be applicable we must reexpress the null geodesics in terms of an affine parameter $\lambda_\ast$. Given a solution $x(\lambda)$
of (\ref{tredue}) we compute the function $\lambda_\ast(\lambda)$ by integrating
\be\label{trecinque}
\frac{d}{d \lambda} \lambda_\ast(\lambda)=\exp \int^{\lambda} d\lambda' \; {\cal K}(x(\lambda'))
\ee
and determine its inverse $\lambda=\lambda(\lambda_\ast)$. Then we define $x^{\mu}_\ast({\lambda_\ast})\equiv
x^{\mu}(\lambda(\lambda_\ast))$ which satisfies
\be\label{3.6}
\frac{d }{d\lambda_\ast} x_\ast^\mu(\lambda_\ast)= n^\mu (x_\ast(\lambda_\ast))
\ee
Here 
\be\label{3.7}
n^\mu (x) = e^{-\Gamma(x)}u^{\mu}(x)
\ee
is a new vector field, with $\Gamma(x)$  satisfying 
\be\label{3.8}
u^\mu \partial_\mu \Gamma ={\cal K}
\ee 
Using (\ref{3.8}) one easily verifies that $n^\nu D_\nu n^\mu=0$, implying that $\lambda_\ast$ is an affine parameter \cite{poi}.

The  expansion scalar $\Theta$ which determines the location of the AH is the divergence of $n^\mu$. Using (\ref{3.8}) we find
\be
\Theta \equiv D_\mu n^\mu = e^{-\Gamma}[D_\mu u^\mu - {\cal K}].
\ee
For $f(r,v)=1-2G(r)M(v)/r$ we have 
\be\label{3.9}
D_\mu u^\mu = \frac{1}{r} f + \frac{1}{2}\partial_r f
\ee
Together with (\ref{trequattro}) this yields the expansion scalar
\be\label{3.10}
\Theta=\frac{1}{r} e^{-\Gamma} f
\ee
This is the same result as for the classical Vaidya metric; the $r$-dependence of $G$ did not lead to extra terms.
 
Eq.(\ref{3.10}) tells us that $\Theta$ vanishes if, and only if, $f$ vanishes. According to point a) of the Introduction
this is the case at $r=r_{+}$ and $r=r_{-}$ with $r_{\pm}$ defined by (\ref{sette}) with (\ref{otto}). (We assume that
the $r$-dependence of $G$ is given by (\ref{quattro}) with $\gamma=0$). Since 
$r_{\pm}$ depends on $M$, it has become a function of the advanced time $v$ now: 
$r_{\pm}(v)\equiv r_{\pm}(M(v))$. Defined as the outermost trapped surface, the AH is characterized by the implicit
equation $r=r_{+}(v)$ where, explicitly,
\be\label{3.11}
r_{+}(v)=\frac{\widetilde\omega}{M_{\rm cr}} \; \Bigg [ \frac{M(v)}{M_{\rm cr}}+ \sqrt{ \Big (\frac{M(v)}{M_{\rm cr}} \Big)^2-1} \Bigg ]
\ee

The second horizon surface, the TLS, is defined  as the locus where the 4-velocity of static observers 
$u^{\alpha} \propto \delta^{\alpha}_\tau$ becomes lightlike, with $\partial / \partial \tau $
a vector orthogonal to the $r=const$ hypersurfaces. Since we consider 
a spherically symmetric spacetime,  this vector is precisely  $\partial / \partial v$ and  AH and TLS
coincide in this case, being  $u^\alpha u^\beta g_{\alpha\beta}=g_{vv}=-f$.

Strictly speaking, apart from the outer AH $r=r_{+}(v)$ there exists also an inner TLS=AH 
$r=r_{-}(v)$, at which the vector field $\partial /\partial v$ switches back from spacelike to timelike. 
\begin{figure}
\includegraphics[width=12cm]{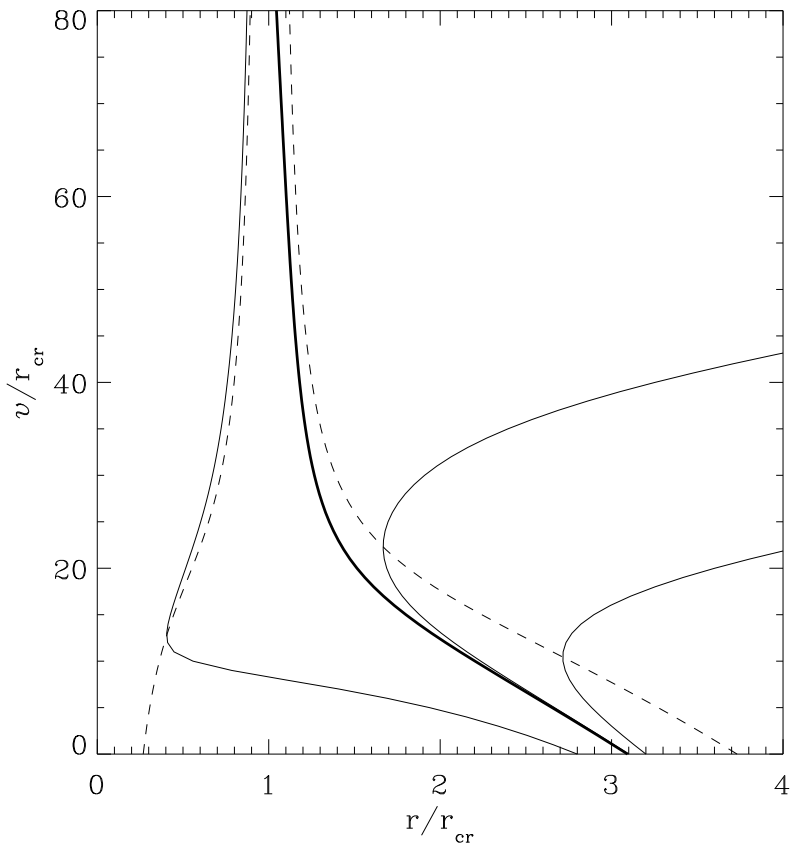}
\caption{\label{fig6}
Light rays of the outgoing null congruence for $M/M_{\rm cr} = 2$. 
The thick solid line is the EH determined numerically, the dashed lines are the outer and inner TLS=AH, 
and the thin solid lines are generic null rays, inside and outside the EH.
}
\end{figure}
\renewcommand{\theequation}{4.\arabic{equation}}
\setcounter{equation}{0}
\section{The event horizon}
The radial light rays $r=r(v)$ of the outgoing null congruence are to be found by solving the differential equation
$\dot{r}(v)=f(r(v),v)/2$, or explicitly, 
\be\label{4.1}
\frac{d r(v)}{dv}  = \frac{1}{2} \Big ( 1-\frac{2 G(r(v)) M(v)}{r(v)}\Big )
\ee
Depending on their initial points, light rays can, or can not, escape to infinity for $v\rightarrow \infty$. 
By definition, the ``separatrix" separating those two classes of solutions is the event horizon, the 
outermost locus traced by outgoing photons that can never reach arbitrarily large distances. 

It is easy to understand why for a radiating (as opposed to an accreting) black hole the EH is {\it inside} the TLS. 
Inside (outside) the TLS, $f$ is negative (positive), implying that the light ray's $r(v)$ decreases (increases) 
if $r(v)<r_{\rm TLS}(v)$ $ \big ( r(v)>r_{\rm TLS}(v) \big )$.  
Under certain conditions it can happen that due to the hole's mass 
loss the radius of the TLS decreases faster than $r(v)$. As a result, 
the light ray intersects the TLS, with $\dot{r}=0$ there, and then escapes from the hole with $\dot{r}>0$. 

The situation is illustrated in Fig.(\ref{fig6}). The region between $r_{\rm EH}$ and $r_{\rm TLS}$ forms the hole's 
``quantum ergosphere" which owes its existence entirely to the evaporation process.

Clearly a determination of the EH's $r_{\rm EH}(v)$ requires 
knowledge of $M(v)$ for arbitrarily late times $v$ even. Therefore the semiclassical
approximation is not sufficient to find the EH since it breaks down for small $M$. 
As a way out, York \cite{ybook,y1} proposed to replace
the above rigorous definition of the EH by an approximate criterion which is local in $v$ and does not require the ``teleological"
information about $M(v)$ at late $v$. It is supposed to be valid for small $L$. In the following we analyze (\ref{4.1})
both using the approximate criterion and the exact definition of the EH. Since we have an explicit prediction for the final stages of the
evaporation process we are in a position to determine the EH exactly, and thus to assess the validity and precision of York's approximate
``working definition".

\begin{figure}
\includegraphics[width=10cm]{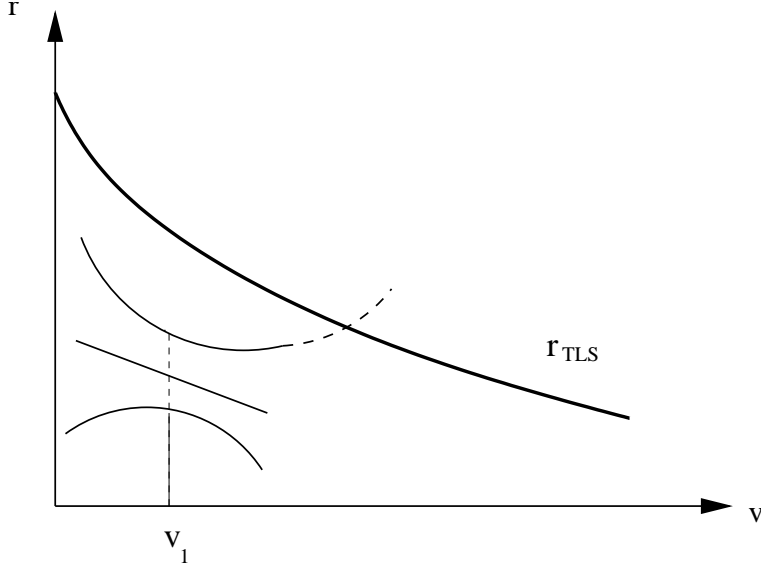}
\caption{\label{yo}
Light rays of the outging null congruence with $\ddot{r}>0$, $\ddot{r}=0$, and $\ddot{r}<0$ at $v_{1}$. Those
with $\ddot{r}>0$ are most likely to eventually cross the TLS. 
}
\end{figure}

Let us begin with this working definition, applied to the RG improved spacetime. While according to Eq.(\ref{4.1}), the ``velocity" $\dot{r}(v)$
is strictly negative inside the TLS, the acceleration $\ddot{r}(v)$ can have either sign there. York proposes to approximately
identify $r_{\rm EH}(v)$ with the radius where the light ray's ``acceleration" vanishes: $\ddot{r}({\rm EH})=0$. 
In pictorial terms 
one could think of this condition as separating the light rays curving downward ($\ddot{r}<0$) from those curving upward $(\ddot{r}>0)$. Clearly
the former (latter) are very unlikely (likely) to subsequently cross the TLS, even if this is not a rigorous criterion, of course.
(See Fig.(\ref{yo}) for a schematic sketch.)
At least in the Hawking regime, photons are imprisoned by the approximate horizon for times long compared to the dynamical time scale
of the evaporating hole \cite{ybook, y1}. 

Taking a second $v$-derivative of (\ref{4.1}) we obtain the ``acceleration" 
\be\label{4.2}
{\ddot r}(v)= L(v) \frac{G(r)}{r} + G(r) M(v)\frac{\dot r}{r^2} -
{G}'(r) M(v) {\dot r \over r}
\ee
Here we assumed that $G(r)$ does not have a parametric dependence on $M$, which is actually true for $\gamma=0$:
\be\label{4.3}
G(r)=\frac{G_0 r^2}{r^2+{\widetilde\omega}G_0}
\ee
(For general $\gamma$ there appears an additional term $\propto dG/dM$ in (\ref{4.2}) which is irrelevant qualitatively.)
Eq.(\ref{4.2}) tells us that, when $\ddot r = 0$, the radius $r$, the velocity $\dot r$, and the time $v$ are related by
\be\label{4.4}
\dot{r}=-L \; G(r) \Big [ \frac{G(r) M(v)}{r} - M(v){G}'(r)\Big ]^{-1}
\ee  
Since $r_{\rm EH}=r_{\rm TLS}$ if the hole would not radiate, and since higher orders in the luminosity are neglected,
the difference $r_{\rm TLS}-r_{\rm EH}$ which we would like to compute is of order $L$. For this reason we may replace
$r\equiv r_{\rm EH}$ on the RHS of (\ref{4.4}) with $r_{\rm TLS}=r_{+}(v)$, the error being of order $L^2$:
\be\label{4.5}
\dot{r}= - 2 L  G(r_{+}) \Big [\frac{2 G(r_{+}) M(v)}{r_{+}}-2 M(v) G'(r_{+}) \Big ]^{-1}
= {-2L G(r_{+}) \over  [ 1-2  M(v)G'(r_{+})]  }
\ee
In the second equality of (\ref{4.5}) we used that $2G(r_{+})M=r_{+}$ which follows from $f(r_{+})=0$.
Eq.(\ref{4.5}) provides us with the ``velocity" $\dot{r}$ at the point where the acceleration vanishes. The corresponding
coordinate, $r_{\rm EH}$, is the approximate location of the EH. We obtain it by using $\dot{r}=f(r)/2$ in order to rewrite
the LHS of (\ref{4.5}) in the form $f(r_{\rm EH})/2$, and then inverting the function $f$. This inversion is easy to perform
since, again, we may expand in $r_{\rm EH}-r_{+}=O(L)$:
\be\label{x}\nonumber
f(r_{\rm EH})=f(r_{+})+(r_{\rm EH}-r_{\rm +})f'(r_{+}) + O(L^2) . 
\ee
Hence, to order $L$, ${\dot r}({\rm EH})=(r_{\rm EH}-r_{+})f'(r_{+})/2$ which yields, together with (\ref{4.5}),
\be\label{4.6}
r_{\rm EH}=r_{+}- \frac{4 L G(r_{+})}{f'(r_{+})[1-2M G'(r_{+})]}
\ee
Differentiating $f(r)=1-2G(r)M/r$ and using $2G(r_{+})M=r_{+}$ one obtains the relation $f'(r_{+})=[1-2 M G'(r_{+})]/r_{+}$.
This leads to the following explicit formula for the $v$-dependence of $r_{\rm EH}$:
\be\label{4.7}
r_{\rm EH}(v)=r_{+}(v)\Bigg \{ 1-\frac{4 \; L(v) \; G(r_{+}(v))}{[1-2\;M(v)\;G'(r_{+}(v))]^2} \Bigg \}
\ee
Eq.(\ref{4.7}) is valid for an arbitrary $r$-dependence of $G$, still. If $G=const$ we recover 
York's result $r_{\rm EH}=r_{+}[1-4G_0L]$ with $r_{+}=2G_0 M$.

\begin{figure}
\includegraphics[width=12cm]{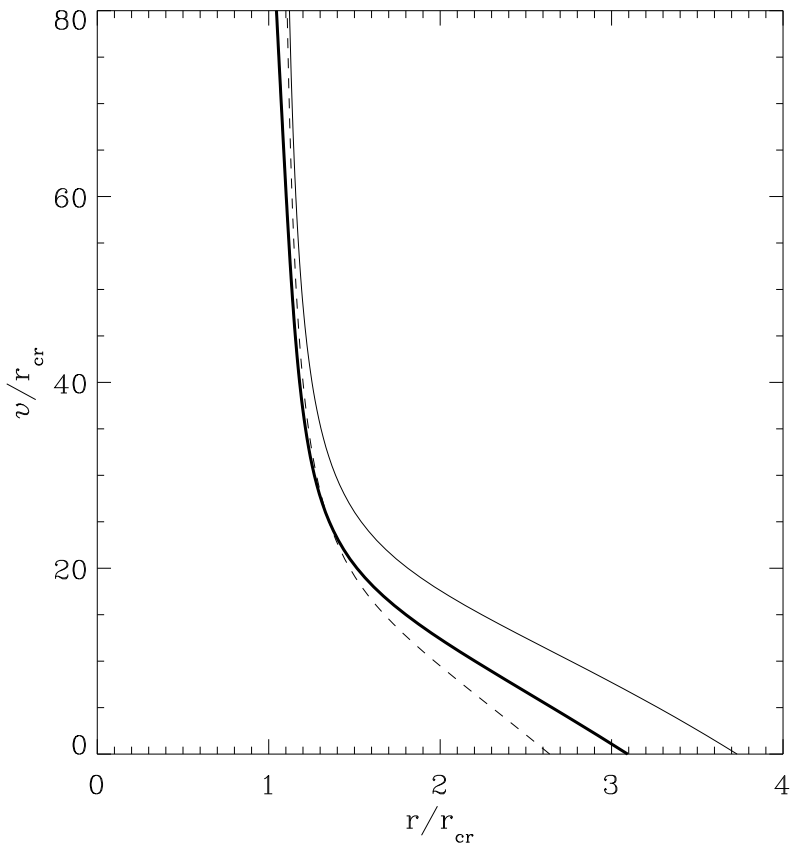}
\caption{\label{yyy}
The EH determined numerically (solid thick line) and by York's approximate  criterion (dashed line)
for $M/M_{\rm cr} = 2$. The solid thin line is the AH.
}
\end{figure}

Let us now specialize for the function $G(r)$ motivated by quantum gravity, Eq.(\ref{4.3}). This $r$-dependence
of Newton's constant implies that, at the TLS,
$G(r_{+})=r_{+}/(2M)$ and $G'(r_{+}) ={\widetilde \omega}/(2M^2 r_{+})$. 
These equations were simplified using the relation $r_{+}^2+\widetilde{\omega} G_0 = 2 G_0 M r_{+}$ which is equivalent to
$f(r_{+})=0$ with (\ref{4.3}). They lead to the following result for the position of the EH:
\begin{subequations}\label{4.8}
\ba\label{4.8a}
&&\re=\rp \big [1-4\;G_0 \; L(M) \; Y(\Omega) \big ]\\[2mm]
&&Y(\Omega)\equiv \frac{1+\sqrt{1-\Omega}}{2(1-\Omega)}
\ea
\end{subequations}
The radius $\rp$ is given by Eq.(\ref{sette}) or (\ref{3.11}), respectively, 
and $M$ and $\Omega\equiv M_{\rm cr}^2/M(v)^2$ are understood to be functions of $v$, of course. The correction factor $Y(\Omega)$
measures the deviations from the semiclassical result; we have $Y(\Omega)\approx 1$ everywhere in the Hawking regime 
($\Omega \approx 0$). For $\Omega \nearrow 1$ the corrections become large; in fact, $Y(\Omega)$ diverges at $\Omega =1$.
Nevertheless, the product $LY$ vanishes for $\Omega\nearrow 1$, $M\searrow M_{\rm cr}$, which becomes clear when we insert
(\ref{dieci}) into (\ref{4.8a}):
\begin{subequations}\label{4.9}
\ba\label{4.9a}
\re&&=\rp \Bigg [ 1-\frac{2\sigma}{(4\pi)^3\widetilde{\omega}}\; \frac{\Omega (1-\Omega)}{[1+\sqrt{1-\Omega}]}\Bigg ]\\[2mm] 
&&=\rp-\frac{2\sigma}{(4\pi)^3}\; \frac{1}{M}\Big (1-\frac{M_{\rm cr}^2}{M^2} \Big )
\ea
\end{subequations}
This is our final result for the radius of the event horizon, as given by York's approximate local criterion. 
In the early stages of the evaporation we recover the semiclassical result. For $v\rightarrow \infty$
however, $M(v)\rightarrow M_{\rm cr}$, and both $r_{\rm EH}$ and $r_{+}(v)$ approach the same limiting value asymptotically:
$r_{\rm cr}\equiv r_{\pm}(M_{\rm cr})= \sqrt{\widetilde{\omega} G_0}=\sqrt{\widetilde{\omega}}\ell_{\rm pl}$. This behavior  can 
be seen in Fig.(\ref{yyy}).  

As we anticipated, we now see that $\re$ is indeed smaller than $\rp$ during the entire evaporation process, the EH is 
{\it inside} the (outer) TLS, thus giving rise to a quantum ergosphere.

It is an important observation that, while $r_{\rm EH}$ is smaller than $\rp$, it is always larger than 
$r_{-}$, provided the radiation effects are small compared to the quantum gravity effects in an appropriate sense.
Comparing the radius (\ref{4.9}) to $r_{-}$ as given by (\ref{sette}) we find that 
$\re (M) > r_{-}(M)$ provided 
\be\label{4.10}
\frac{\sigma}{(4\pi)^3\widetilde{\omega}}< \frac{1}{\Omega{\sqrt{1-\Omega}}}
\ee
For $0\leq \Omega \leq 1$ the RHS of (\ref{4.10}) is bounded below by the constant $3\sqrt{3}/2$. As a result, 
$\re$ is larger than $r_{-}$ during the entire evaporation process provided
\be\label{4.11}
\sigma < \frac{3}{2}\sqrt{3} (4\pi)^3 \; \widetilde{\omega}
\ee
If (\ref{4.11}) is satisfied, we have $\rp(v)>\re (v) >r_{-} (v)$ for any finite $v$, and the EH touches both the outer and the inner TLS only
in the limit $v\rightarrow \infty$ where $\rp,\re,r_{-}\rightarrow r_{\rm cr}$. 

If (\ref{4.11}) is violated our method is inapplicable, most probably, and the improved Vaidya metric is not a reliable 
description of the spacetime structure. This metric is valid to first order in $L$ only, which means that the dimensionless luminosity
has to be much smaller than unity,  $G_0 L \ll 1$. If so, Eq.(\ref{dieci}) implies $\sigma/[(4\pi)^3\widetilde{\omega}]\ll 1$,
and (\ref{4.11}) is indeed satisfied. (Note that $M_{\rm cr}^2/\widetilde{\omega}^2= G_0^{-1}/\widetilde{\omega}$.) It is 
reassuring to see that, for pure gravity, Eq.(\ref{4.11}) is fulfilled with a very broad margin; the expansion parameter 
$\sigma/[(4\pi)^3\widetilde{\omega}]$ assumes the tiny value $8.3 \cdot 10^{-5} / \tilde{\omega}$ 
in this case, with $\tilde\omega = O(1)$.

Let us return to the exact definition of the EH now. We compute it for our model by numerically solving (\ref{4.1}) for a set of 
initial conditions $r(0)$. For a certain range of $r(0)$'s the trajectories will ultimately cross the TLS and escape to infinity, 
while the others remain at radii below 
$r_{\rm TLS}$ for arbitrarily late advanced times. The light ray separating those two classes of solutions defines the EH, 
a null hypersurface by construction. 

In Fig.(\ref{fig6}) we show various trajectories of either class, as well as the EH, 
determined numerically from its {\it exact} definition. We have numerically evolved   
several initial conditions $r(0)$ in
order to determine the boundary between trapped and escaping null geodesics in the limit $v\rightarrow \infty$.
\begin{figure}
\includegraphics[width=6cm]{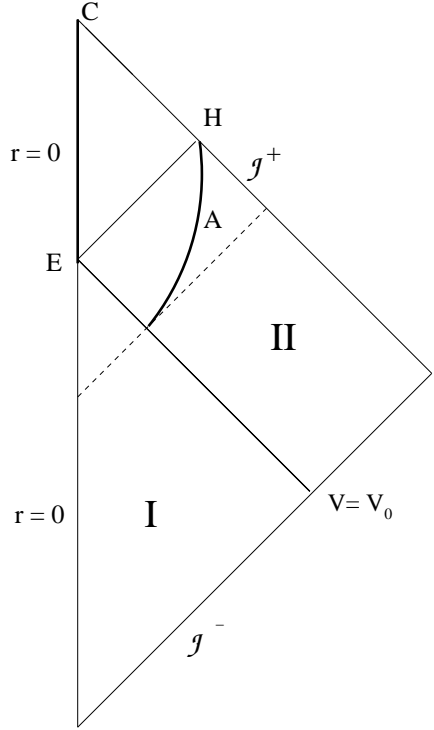}
\caption{\label{fig7}
The conformal diagram of the evaporating quantum black hole:   
region I is  a flat spacetime, and 
region II is the evaporating BH spacetime, 
$EH$ is the event horizon, $CH$ is the inner (Cauchy) horizon,
and $A$ is the apparent horizon. }
\end{figure}
In Fig.(\ref{yyy}) the true EH is compared to the prediction of York's criterion. Obviously the
latter provides a rather accurate approximation to the true horizon.

The global structure of the spacetime is depicted in the conformal diagram in Fig.(\ref{fig7}).
Region I is a flat spacetime, while at $V=V_0$ ($V$ is the Kruskal advanced time coordinate, 
defined as $V =-\exp (-\kappa v)$ being $\kappa$ the surface gravity of the outer horizon)  
an imploding null shell is present (strictly speaking it must have a negative tension  
in order to balance the flux of negative energy on its future side \cite{baris}).  
Region II is the evaporating black hole spacetime.  The AH is a timelike hypersurface which 
``meets'' the EH at  future null infinity in the conformal diagram. 
The null ray which is tangent to the earliest portion of the apparent horizon 
$A$ would have been the EH if the hole were not radiating.  
The final state of the black hole is an extremal black hole whose inner 
and outer horizons have the same radius $(r=r_{\rm cr})$ and are located  at 
the event horizon $EH$ and the inner (Cauchy) horizon $CH$ in  Fig.(\ref{fig7}).   

It is instructive to compare the areas ${\cal A}$ of the various horizons. They are defined by intersecting the EH, AH, and TLS with the 
incoming null surfaces $v= const$. Thus ${\cal A}_{\rm TLS} (v)= 4\pi \rp(v)^2$ and ${\cal A}_{\rm EH}(v)=4\pi\re(v)^2$.
From Eq.(\ref{sette})
we obtain for ${\cal A}_{\rm TLS} \equiv {\cal A}_{\rm AH}$
\be\label{4.12}
{\cal A}_{\rm TLS} = 4\pi G_0^2 M^2 \Big [1+\sqrt{1-(M_{\rm cr}/M)^2}\Big ]^2
\ee 
and the approximate result (\ref{4.9a}) for the  event horizon implies 
\be\label{4.13}
{\cal A}_{\rm EH}={\cal A}_{\rm TLS} \Big [1-\frac{4\sigma}{(4\pi)^3\widetilde{\omega}}
\frac{\Omega(1-\Omega)}{1+\sqrt{1-\Omega}} \Big ]
\ee
where a term of second order in $\sigma/(4\pi)^3\widetilde{\omega}$ has been neglected. The difference 
$\delta {\cal A}\equiv {\cal A}_{\rm TLS}-{\cal A}_{\rm EH}$ is
given by
\be\label{4.14}
\delta {\cal A}=\frac{\sigma}{4\pi^2} \; \ell_{\rm Pl}^2 \; (1-\Omega) \big [1+\sqrt{1-{\Omega}} \big ]
\ee
During the early stages of the evaporation process, 
$\delta {\cal A}\approx \sigma l_{\rm Pl}^2 /(2\pi^2) = 128 \pi B \ell_{\rm Pl}^2$ which coincides with the known result
\cite{ybook} for the Hawking regime, while $\delta {\cal A}$ vanishes proportional to $(M^2-M^2_{\rm cr})\rightarrow 0$
for $v \rightarrow \infty$.
It had been emphasized by York \cite{ybook,y1} that in the Hawking regime he considered, $\delta {\cal A}$ is a universal
({\it i.e.} $M$ independent) quantity which depends only on $\sigma$, thus counting the degrees of freedom of the 
field quanta which can be evaporated off. Looking at Eq.(\ref{4.14}) we see that this universality does not 
persist beyond the semiclassical approximation.
\renewcommand{\theequation}{5.\arabic{equation}}
\setcounter{equation}{0}
\section{Conclusion}
The renormalization group improvement of  black hole spacetimes
according to Quantum Einstein Gravity  leads to concrete predictions on the final state of 
the evaporation process.
Unlike previous studies based on {\it ad hoc} modifications
of the equation of state of matter at very high (Planckian) densities \cite{muka,dymni1,hay05},
or models based on loop quantum gravity \cite{ashte05},
the mass of the remnant can be calculated explicitly: 
$M_{\rm cr}=\sqrt{\widetilde{\omega}} \ell_{\rm Pl}$. 
Its precise value is  determined by the value of $\widetilde{\omega}$
which is a measurable quantity in principle 
\cite{donog}. 
No naked singularity forms, at variance with the paradigm proposed in \cite{his,hay05}, so that   
the remnant is a mini-black hole of Planckian size. 
On the other hand, it is intriguing to note that exactly solvable semiclassical gravity-dilaton models
predict a final state described by an extremal configuration which is reached in an infinite amount 
of time \cite{fabbri}. (See also \cite{grumi} and   
\cite{jac} for analogous semiclassical  models, and 
\cite{ward1,ward2} for  an approach based on special resummations of higher order graviton
loops. )

It would be interesting 
to investigate the possible astrophysical implications of a population  of
stable Planck size mini-black holes produced in the Early Universe or by the interaction of 
cosmic rays with the interstellar medium \cite{barrau05,barrau03}. We hope to address these issues in 
a subsequent publication.

\section{Acknowledgments}
We would like to thank Werner Israel for helpful suggestions on an earlier version of this manuscript. 
A.B. would also like to thank the Department of Physics of Mainz University  and M.R. 
the Catania Astrophysical Observatory for the financial support and for the cordial hospitality 
extended to them while this work was in progress.  
We are also grateful to INFN, Sezione di Catania for financial support.

\bibliography{bb.bbl}

\begin{thebibliography}{60}
\expandafter\ifx\csname natexlab\endcsname\relax\def\natexlab#1{#1}\fi
\expandafter\ifx\csname bibnamefont\endcsname\relax
  \def\bibnamefont#1{#1}\fi
\expandafter\ifx\csname bibfnamefont\endcsname\relax
  \def\bibfnamefont#1{#1}\fi
\expandafter\ifx\csname citenamefont\endcsname\relax
  \def\citenamefont#1{#1}\fi
\expandafter\ifx\csname url\endcsname\relax
  \def\url#1{\texttt{#1}}\fi
\expandafter\ifx\csname urlprefix\endcsname\relax\def\urlprefix{URL }\fi
\providecommand{\bibinfo}[2]{#2}
\providecommand{\eprint}[2][]{\url{#2}}

\bibitem[{\citenamefont{{Hawking}}(1975)}]{hawking}
\bibinfo{author}{\bibfnamefont{S.}~\bibnamefont{{Hawking}}},
  \bibinfo{journal}{Commun. Math. Phys.} \textbf{\bibinfo{volume}{43}},
  \bibinfo{pages}{199} (\bibinfo{year}{1975}).

\bibitem[{\citenamefont{York}(1984)}]{ybook}
\bibinfo{author}{\bibfnamefont{J.}~\bibnamefont{York}}, in
  \emph{\bibinfo{booktitle}{Quantum Theory of Gravity: Essays in Honor of the
  Sixtieth Birthday of \\ Bryce S. DeWitt}}, edited by
  \bibinfo{editor}{\bibfnamefont{S.}~\bibnamefont{Christensen}}
  (\bibinfo{publisher}{Adam Hilger}, \bibinfo{address}{Bristol},
  \bibinfo{year}{1984}).

\bibitem[{\citenamefont{{York}}(1983)}]{y1}
\bibinfo{author}{\bibfnamefont{J.}~\bibnamefont{{York}}},
  \bibinfo{journal}{\prd} \textbf{\bibinfo{volume}{28}}, \bibinfo{pages}{2929}
  (\bibinfo{year}{1983}).

\bibitem[{\citenamefont{{York}}(1985)}]{y2}
\bibinfo{author}{\bibfnamefont{J.}~\bibnamefont{{York}}},
  \bibinfo{journal}{\prd} \textbf{\bibinfo{volume}{31}}, \bibinfo{pages}{775}
  (\bibinfo{year}{1985}).

\bibitem[{\citenamefont{{Hajicek} and {Israel}}(1980)}]{is}
\bibinfo{author}{\bibfnamefont{P.}~\bibnamefont{{Hajicek}}} \bibnamefont{and}
  \bibinfo{author}{\bibfnamefont{W.}~\bibnamefont{{Israel}}},
  \bibinfo{journal}{Phys. Lett.} \textbf{\bibinfo{volume}{A80}},
  \bibinfo{pages}{9} (\bibinfo{year}{1980}).

\bibitem[{\citenamefont{Poisson}(2004)}]{poi}
\bibinfo{author}{\bibfnamefont{E.}~\bibnamefont{Poisson}},
  \emph{\bibinfo{title}{A Relativist's Toolkit}} (\bibinfo{publisher}{Cambridge
  University Press}, \bibinfo{year}{2004}).

\bibitem[{\citenamefont{{Reuter}}(1998)}]{mr}
\bibinfo{author}{\bibfnamefont{M.}~\bibnamefont{{Reuter}}},
  \bibinfo{journal}{\prd} \textbf{\bibinfo{volume}{57}}, \bibinfo{pages}{971}
  (\bibinfo{year}{1998}).

\bibitem[{\citenamefont{Dou and Percacci}(1998)}]{percadou}
\bibinfo{author}{\bibfnamefont{D.}~\bibnamefont{Dou}} \bibnamefont{and}
  \bibinfo{author}{\bibfnamefont{R.}~\bibnamefont{Percacci}},
  \bibinfo{journal}{Class. Quantum Grav.} \textbf{\bibinfo{volume}{15}},
  \bibinfo{pages}{3449} (\bibinfo{year}{1998}).

\bibitem[{\citenamefont{{Lauscher} and {Reuter}}(2002{\natexlab{a}})}]{oliver1}
\bibinfo{author}{\bibfnamefont{O.}~\bibnamefont{{Lauscher}}} \bibnamefont{and}
  \bibinfo{author}{\bibfnamefont{M.}~\bibnamefont{{Reuter}}},
  \bibinfo{journal}{\prd} \textbf{\bibinfo{volume}{65}},
  \bibinfo{pages}{025013} (\bibinfo{year}{2002}{\natexlab{a}}).

\bibitem[{\citenamefont{{Reuter} and
  {Saueressig}}(2002{\natexlab{a}})}]{frank1}
\bibinfo{author}{\bibfnamefont{M.}~\bibnamefont{{Reuter}}} \bibnamefont{and}
  \bibinfo{author}{\bibfnamefont{F.}~\bibnamefont{{Saueressig}}},
  \bibinfo{journal}{\prd} \textbf{\bibinfo{volume}{65}},
  \bibinfo{pages}{065016} (\bibinfo{year}{2002}{\natexlab{a}}).

\bibitem[{\citenamefont{{Lauscher} and {Reuter}}(2002{\natexlab{b}})}]{oliver2}
\bibinfo{author}{\bibfnamefont{O.}~\bibnamefont{{Lauscher}}} \bibnamefont{and}
  \bibinfo{author}{\bibfnamefont{M.}~\bibnamefont{{Reuter}}},
  \bibinfo{journal}{Class. Quantum Grav.} \textbf{\bibinfo{volume}{19}},
  \bibinfo{pages}{483} (\bibinfo{year}{2002}{\natexlab{b}}).

\bibitem[{\citenamefont{Souma}(1999)}]{souma}
\bibinfo{author}{\bibfnamefont{W.}~\bibnamefont{Souma}},
  \bibinfo{journal}{Prog. Theor. Phys.} \textbf{\bibinfo{volume}{102}},
  \bibinfo{pages}{181} (\bibinfo{year}{1999}).

\bibitem[{\citenamefont{{Percacci} and
  {Perini}}(2003{\natexlab{a}})}]{percacciperini}
\bibinfo{author}{\bibfnamefont{R.}~\bibnamefont{{Percacci}}} \bibnamefont{and}
  \bibinfo{author}{\bibfnamefont{D.}~\bibnamefont{{Perini}}},
  \bibinfo{journal}{\prd} \textbf{\bibinfo{volume}{68}},
  \bibinfo{pages}{044018} (\bibinfo{year}{2003}{\natexlab{a}}).

\bibitem[{\citenamefont{{Reuter} and
  {Saueressig}}(2002{\natexlab{b}})}]{frank2}
\bibinfo{author}{\bibfnamefont{M.}~\bibnamefont{{Reuter}}} \bibnamefont{and}
  \bibinfo{author}{\bibfnamefont{F.}~\bibnamefont{{Saueressig}}},
  \bibinfo{journal}{\prd} \textbf{\bibinfo{volume}{66}},
  \bibinfo{pages}{125001} (\bibinfo{year}{2002}{\natexlab{b}}).

\bibitem[{\citenamefont{Litim}(2004)}]{litimgrav}
\bibinfo{author}{\bibfnamefont{D.~F.} \bibnamefont{Litim}},
  \bibinfo{journal}{Phys. Rev. Lett.} \textbf{\bibinfo{volume}{92}},
  \bibinfo{pages}{201301} (\bibinfo{year}{2004}).

\bibitem[{\citenamefont{Niedermaier}(2003)}]{max}
\bibinfo{author}{\bibfnamefont{M.}~\bibnamefont{Niedermaier}},
  \bibinfo{journal}{Nucl. Phys.} \textbf{\bibinfo{volume}{B673}},
  \bibinfo{pages}{131} (\bibinfo{year}{2003}).

\bibitem[{\citenamefont{{Lauscher} and {Reuter}}(2002{\natexlab{c}})}]{oliver3}
\bibinfo{author}{\bibfnamefont{O.}~\bibnamefont{{Lauscher}}} \bibnamefont{and}
  \bibinfo{author}{\bibfnamefont{M.}~\bibnamefont{{Reuter}}},
  \bibinfo{journal}{\prd} \textbf{\bibinfo{volume}{66}},
  \bibinfo{pages}{025026} (\bibinfo{year}{2002}{\natexlab{c}}).

\bibitem[{\citenamefont{{Percacci} and
  {Perini}}(2003{\natexlab{b}})}]{percaper2}
\bibinfo{author}{\bibfnamefont{R.}~\bibnamefont{{Percacci}}} \bibnamefont{and}
  \bibinfo{author}{\bibfnamefont{D.}~\bibnamefont{{Perini}}},
  \bibinfo{journal}{\prd} \textbf{\bibinfo{volume}{67}},
  \bibinfo{pages}{081503} (\bibinfo{year}{2003}{\natexlab{b}}).

\bibitem[{\citenamefont{Reuter and Saueressig}(2004)}]{frankf}
\bibinfo{author}{\bibfnamefont{M.}~\bibnamefont{Reuter}} \bibnamefont{and}
  \bibinfo{author}{\bibfnamefont{F.}~\bibnamefont{Saueressig}},
  \bibinfo{journal}{Fortsch. Phys.} \textbf{\bibinfo{volume}{52}},
  \bibinfo{pages}{650} (\bibinfo{year}{2004}).

\bibitem[{\citenamefont{Niedermaier}(2002)}]{max2}
\bibinfo{author}{\bibfnamefont{M.}~\bibnamefont{Niedermaier}},
  \bibinfo{journal}{JHEP} \textbf{\bibinfo{volume}{12}}, \bibinfo{pages}{066}
  (\bibinfo{year}{2002}).

\bibitem[{\citenamefont{Forgacs and Niedermaier}(2002)}]{max3}
\bibinfo{author}{\bibfnamefont{P.}~\bibnamefont{Forgacs}} \bibnamefont{and}
  \bibinfo{author}{\bibfnamefont{M.}~\bibnamefont{Niedermaier}}
  (\bibinfo{year}{2002}), \eprint{hep-th/0207028}.

\bibitem[{\citenamefont{Lauscher and Reuter}(2002)}]{oliver4}
\bibinfo{author}{\bibfnamefont{O.}~\bibnamefont{Lauscher}} \bibnamefont{and}
  \bibinfo{author}{\bibfnamefont{M.}~\bibnamefont{Reuter}},
  \bibinfo{journal}{Int. J. Mod. Phys. A} \textbf{\bibinfo{volume}{17}},
  \bibinfo{pages}{993} (\bibinfo{year}{2002}).

\bibitem[{\citenamefont{Lauscher and Reuter}(2005{\natexlab{a}})}]{oliver5}
\bibinfo{author}{\bibfnamefont{O.}~\bibnamefont{Lauscher}} \bibnamefont{and}
  \bibinfo{author}{\bibfnamefont{M.}~\bibnamefont{Reuter}},
  \bibinfo{journal}{JHEP} \textbf{\bibinfo{volume}{10}}, \bibinfo{pages}{050}
  (\bibinfo{year}{2005}{\natexlab{a}}).

\bibitem[{\citenamefont{Lauscher and Reuter}(2005{\natexlab{b}})}]{oliver6}
\bibinfo{author}{\bibfnamefont{O.}~\bibnamefont{Lauscher}} \bibnamefont{and}
  \bibinfo{author}{\bibfnamefont{M.}~\bibnamefont{Reuter}}
  (\bibinfo{year}{2005}{\natexlab{b}}), \eprint{hep-th/0511260}.

\bibitem[{\citenamefont{Reuter and Schwindt}(2006)}]{resh}
\bibinfo{author}{\bibfnamefont{M.}~\bibnamefont{Reuter}} \bibnamefont{and}
  \bibinfo{author}{\bibfnamefont{J.-M.} \bibnamefont{Schwindt}},
  \bibinfo{journal}{JHEP} \textbf{\bibinfo{volume}{0601}}, \bibinfo{pages}{070}
  (\bibinfo{year}{2006}).

\bibitem[{\citenamefont{Bonanno and Reuter}(2005)}]{brproper}
\bibinfo{author}{\bibfnamefont{A.}~\bibnamefont{Bonanno}} \bibnamefont{and}
  \bibinfo{author}{\bibfnamefont{M.}~\bibnamefont{Reuter}},
  \bibinfo{journal}{JHEP} \textbf{\bibinfo{volume}{0502}}, \bibinfo{pages}{035}
  (\bibinfo{year}{2005}).

\bibitem[{\citenamefont{{Bonanno} and {Reuter}}(1999)}]{bh1}
\bibinfo{author}{\bibfnamefont{A.}~\bibnamefont{{Bonanno}}} \bibnamefont{and}
  \bibinfo{author}{\bibfnamefont{M.}~\bibnamefont{{Reuter}}},
  \bibinfo{journal}{\prd} \textbf{\bibinfo{volume}{60}},
  \bibinfo{pages}{084011} (\bibinfo{year}{1999}).

\bibitem[{\citenamefont{{Bonanno} and {Reuter}}(2000)}]{bh2}
\bibinfo{author}{\bibfnamefont{A.}~\bibnamefont{{Bonanno}}} \bibnamefont{and}
  \bibinfo{author}{\bibfnamefont{M.}~\bibnamefont{{Reuter}}},
  \bibinfo{journal}{\prd} \textbf{\bibinfo{volume}{62}},
  \bibinfo{pages}{043008} (\bibinfo{year}{2000}).

\bibitem[{\citenamefont{Bonanno and Reuter}(2002{\natexlab{a}})}]{cosmo1}
\bibinfo{author}{\bibfnamefont{A.}~\bibnamefont{Bonanno}} \bibnamefont{and}
  \bibinfo{author}{\bibfnamefont{M.}~\bibnamefont{Reuter}},
  \bibinfo{journal}{Phys. Rev. D} \textbf{\bibinfo{volume}{65}},
  \bibinfo{pages}{043508} (\bibinfo{year}{2002}{\natexlab{a}}).

\bibitem[{\citenamefont{Bonanno and Reuter}(2002{\natexlab{b}})}]{cosmo2}
\bibinfo{author}{\bibfnamefont{A.}~\bibnamefont{Bonanno}} \bibnamefont{and}
  \bibinfo{author}{\bibfnamefont{M.}~\bibnamefont{Reuter}},
  \bibinfo{journal}{Phys. Lett.} \textbf{\bibinfo{volume}{B527}},
  \bibinfo{pages}{9} (\bibinfo{year}{2002}{\natexlab{b}}).

\bibitem[{\citenamefont{Bentivegna et~al.}(2004)\citenamefont{Bentivegna,
  Bonanno, and Reuter}}]{elo}
\bibinfo{author}{\bibfnamefont{E.}~\bibnamefont{Bentivegna}},
  \bibinfo{author}{\bibfnamefont{A.}~\bibnamefont{Bonanno}}, \bibnamefont{and}
  \bibinfo{author}{\bibfnamefont{M.}~\bibnamefont{Reuter}},
  \bibinfo{journal}{JCAP} \textbf{\bibinfo{volume}{0401}}, \bibinfo{pages}{001}
  (\bibinfo{year}{2004}).

\bibitem[{\citenamefont{Bonanno et~al.}(2003)\citenamefont{Bonanno, Esposito,
  and Rubano}}]{esposito}
\bibinfo{author}{\bibfnamefont{A.}~\bibnamefont{Bonanno}},
  \bibinfo{author}{\bibfnamefont{G.}~\bibnamefont{Esposito}}, \bibnamefont{and}
  \bibinfo{author}{\bibfnamefont{C.}~\bibnamefont{Rubano}},
  \bibinfo{journal}{Gen. Rel. Grav.} \textbf{\bibinfo{volume}{35}},
  \bibinfo{pages}{1899} (\bibinfo{year}{2003}).

\bibitem[{\citenamefont{Tsuneyama}(2004)}]{scalfact}
\bibinfo{author}{\bibfnamefont{T.}~\bibnamefont{Tsuneyama}}
  (\bibinfo{year}{2004}), \eprint{hep-th/0401110}.

\bibitem[{\citenamefont{Reuter and Weyer}(2004)}]{h1}
\bibinfo{author}{\bibfnamefont{M.}~\bibnamefont{Reuter}} \bibnamefont{and}
  \bibinfo{author}{\bibfnamefont{H.}~\bibnamefont{Weyer}},
  \bibinfo{journal}{Phys. Rev. D} \textbf{\bibinfo{volume}{69}},
  \bibinfo{pages}{104022} (\bibinfo{year}{2004}).

\bibitem[{\citenamefont{{Reuter} and {Weyer}}(2004{\natexlab{a}})}]{h2}
\bibinfo{author}{\bibfnamefont{M.}~\bibnamefont{{Reuter}}} \bibnamefont{and}
  \bibinfo{author}{\bibfnamefont{H.}~\bibnamefont{{Weyer}}},
  \bibinfo{journal}{JCAP} \textbf{\bibinfo{volume}{12}}, \bibinfo{pages}{1}
  (\bibinfo{year}{2004}{\natexlab{a}}).

\bibitem[{\citenamefont{{Reuter} and {Weyer}}(2004{\natexlab{b}})}]{wey70}
\bibinfo{author}{\bibfnamefont{M.}~\bibnamefont{{Reuter}}} \bibnamefont{and}
  \bibinfo{author}{\bibfnamefont{H.}~\bibnamefont{{Weyer}}},
  \bibinfo{journal}{\prd} \textbf{\bibinfo{volume}{70}},
  \bibinfo{pages}{124028} (\bibinfo{year}{2004}{\natexlab{b}}).

\bibitem[{\citenamefont{{Bauer}}(2005)}]{bauer}
\bibinfo{author}{\bibfnamefont{F.}~\bibnamefont{{Bauer}}},
  \bibinfo{journal}{Class. Quantum Grav.} \textbf{\bibinfo{volume}{22}},
  \bibinfo{pages}{3533} (\bibinfo{year}{2005}).

\bibitem[{\citenamefont{Wetterich}(1993)}]{avact}
\bibinfo{author}{\bibfnamefont{C.}~\bibnamefont{Wetterich}},
  \bibinfo{journal}{Phys. Lett.} \textbf{\bibinfo{volume}{B301}},
  \bibinfo{pages}{90} (\bibinfo{year}{1993}).

\bibitem[{\citenamefont{Reuter and Wetterich}(1994{\natexlab{a}})}]{ym1}
\bibinfo{author}{\bibfnamefont{M.}~\bibnamefont{Reuter}} \bibnamefont{and}
  \bibinfo{author}{\bibfnamefont{C.}~\bibnamefont{Wetterich}},
  \bibinfo{journal}{Nucl. Phys.} \textbf{\bibinfo{volume}{B417}},
  \bibinfo{pages}{181} (\bibinfo{year}{1994}{\natexlab{a}}).

\bibitem[{\citenamefont{Reuter and Wetterich}(1994{\natexlab{b}})}]{ym2}
\bibinfo{author}{\bibfnamefont{M.}~\bibnamefont{Reuter}} \bibnamefont{and}
  \bibinfo{author}{\bibfnamefont{C.}~\bibnamefont{Wetterich}},
  \bibinfo{journal}{Nucl. Phys.} \textbf{\bibinfo{volume}{B427}},
  \bibinfo{pages}{291} (\bibinfo{year}{1994}{\natexlab{b}}).

\bibitem[{\citenamefont{Reuter and Wetterich}(1993{\natexlab{a}})}]{ym3}
\bibinfo{author}{\bibfnamefont{M.}~\bibnamefont{Reuter}} \bibnamefont{and}
  \bibinfo{author}{\bibfnamefont{C.}~\bibnamefont{Wetterich}},
  \bibinfo{journal}{Nucl. Phys.} \textbf{\bibinfo{volume}{B391}},
  \bibinfo{pages}{147} (\bibinfo{year}{1993}{\natexlab{a}}).

\bibitem[{\citenamefont{Reuter and Wetterich}(1993{\natexlab{b}})}]{ym4}
\bibinfo{author}{\bibfnamefont{M.}~\bibnamefont{Reuter}} \bibnamefont{and}
  \bibinfo{author}{\bibfnamefont{C.}~\bibnamefont{Wetterich}},
  \bibinfo{journal}{Nucl. Phys.} \textbf{\bibinfo{volume}{B408}},
  \bibinfo{pages}{91} (\bibinfo{year}{1993}{\natexlab{b}}).

\bibitem[{\citenamefont{{Bardeen}}(1981)}]{bar}
\bibinfo{author}{\bibfnamefont{J.}~\bibnamefont{{Bardeen}}},
  \bibinfo{journal}{\prl} \textbf{\bibinfo{volume}{23}}, \bibinfo{pages}{2823}
  (\bibinfo{year}{1981}).

\bibitem[{\citenamefont{{Hiscock}}(1981{\natexlab{a}})}]{his}
\bibinfo{author}{\bibfnamefont{W.}~\bibnamefont{{Hiscock}}},
  \bibinfo{journal}{\prd} \textbf{\bibinfo{volume}{23}}, \bibinfo{pages}{2813}
  (\bibinfo{year}{1981}{\natexlab{a}}).

\bibitem[{\citenamefont{{Hiscock}}(1981{\natexlab{b}})}]{his2}
\bibinfo{author}{\bibfnamefont{W.}~\bibnamefont{{Hiscock}}},
  \bibinfo{journal}{\prd} \textbf{\bibinfo{volume}{23}}, \bibinfo{pages}{2823}
  (\bibinfo{year}{1981}{\natexlab{b}}).

\bibitem[{\citenamefont{{Birrell} and {Davies}}(1982)}]{bd}
\bibinfo{author}{\bibfnamefont{N.}~\bibnamefont{{Birrell}}} \bibnamefont{and}
  \bibinfo{author}{\bibfnamefont{P.}~\bibnamefont{{Davies}}},
  \emph{\bibinfo{title}{Quantum Fields in Curved Space}}
  (\bibinfo{publisher}{Cambridge University Press}, \bibinfo{year}{1982}).

\bibitem[{\citenamefont{{Hawking} and {Ellis}}(1973)}]{he}
\bibinfo{author}{\bibfnamefont{S.}~\bibnamefont{{Hawking}}} \bibnamefont{and}
  \bibinfo{author}{\bibfnamefont{G.}~\bibnamefont{{Ellis}}},
  \emph{\bibinfo{title}{The Large Scale Structure of Space-Time}}
  (\bibinfo{publisher}{Cambridge University Press}, \bibinfo{year}{1973}).

\bibitem[{\citenamefont{{Barrab{\`e}s} and {Israel}}(1991)}]{baris}
\bibinfo{author}{\bibfnamefont{C.}~\bibnamefont{{Barrab{\`e}s}}}
  \bibnamefont{and} \bibinfo{author}{\bibfnamefont{W.}~\bibnamefont{{Israel}}},
  \bibinfo{journal}{\prd} \textbf{\bibinfo{volume}{43}}, \bibinfo{pages}{1129}
  (\bibinfo{year}{1991}).

\bibitem[{\citenamefont{Frolov et~al.}(1990)\citenamefont{Frolov, Markov, and
  Mukhanov}}]{muka}
\bibinfo{author}{\bibfnamefont{V.~P.} \bibnamefont{Frolov}},
  \bibinfo{author}{\bibfnamefont{M.~A.} \bibnamefont{Markov}},
  \bibnamefont{and} \bibinfo{author}{\bibfnamefont{V.~F.}
  \bibnamefont{Mukhanov}}, \bibinfo{journal}{Phys. Rev.}
  \textbf{\bibinfo{volume}{D41}}, \bibinfo{pages}{383} (\bibinfo{year}{1990}).

\bibitem[{\citenamefont{Dymnikova}(2003)}]{dymni1}
\bibinfo{author}{\bibfnamefont{I.}~\bibnamefont{Dymnikova}},
  \bibinfo{journal}{Int. J. Mod. Phys.} \textbf{\bibinfo{volume}{D12}},
  \bibinfo{pages}{1015} (\bibinfo{year}{2003}).

\bibitem[{\citenamefont{Hayward}(2005)}]{hay05}
\bibinfo{author}{\bibfnamefont{S.~A.} \bibnamefont{Hayward}}
  (\bibinfo{year}{2005}), \eprint{gr-qc/0506126}.

\bibitem[{\citenamefont{Ashtekar and Bojowald}(2005)}]{ashte05}
\bibinfo{author}{\bibfnamefont{A.}~\bibnamefont{Ashtekar}} \bibnamefont{and}
  \bibinfo{author}{\bibfnamefont{M.}~\bibnamefont{Bojowald}},
  \bibinfo{journal}{Class. Quant. Grav.} \textbf{\bibinfo{volume}{22}},
  \bibinfo{pages}{3349} (\bibinfo{year}{2005}).

\bibitem[{\citenamefont{Donoghue}(1994)}]{donog}
\bibinfo{author}{\bibfnamefont{J.~F.} \bibnamefont{Donoghue}},
  \bibinfo{journal}{Phys. Rev. Lett.} \textbf{\bibinfo{volume}{72}},
  \bibinfo{pages}{2996} (\bibinfo{year}{1994}).

\bibitem[{\citenamefont{Fabbri et~al.}(2000)\citenamefont{Fabbri, Navarro, and
  Navarro-Salas}}]{fabbri}
\bibinfo{author}{\bibfnamefont{A.}~\bibnamefont{Fabbri}},
  \bibinfo{author}{\bibfnamefont{D.~J.} \bibnamefont{Navarro}},
  \bibnamefont{and}
  \bibinfo{author}{\bibfnamefont{J.}~\bibnamefont{Navarro-Salas}},
  \bibinfo{journal}{Phys. Rev. Lett.} \textbf{\bibinfo{volume}{85}},
  \bibinfo{pages}{2434} (\bibinfo{year}{2000}).

\bibitem[{\citenamefont{Grumiller}(2004)}]{grumi}
\bibinfo{author}{\bibfnamefont{D.}~\bibnamefont{Grumiller}},
  \bibinfo{journal}{JCAP} \textbf{\bibinfo{volume}{0405}}, \bibinfo{pages}{005}
  (\bibinfo{year}{2004}).

\bibitem[{\citenamefont{Jacobson}(1998)}]{jac}
\bibinfo{author}{\bibfnamefont{T.}~\bibnamefont{Jacobson}},
  \bibinfo{journal}{Phys. Rev. D} \textbf{\bibinfo{volume}{57}},
  \bibinfo{pages}{4890} (\bibinfo{year}{1998}).

\bibitem[{\citenamefont{Ward}(2005{\natexlab{a}})}]{ward1}
\bibinfo{author}{\bibfnamefont{B.~F.~L.} \bibnamefont{Ward}},
  \bibinfo{journal}{Int. J. Mod. Phys.} \textbf{\bibinfo{volume}{A20}},
  \bibinfo{pages}{3502} (\bibinfo{year}{2005}{\natexlab{a}}).

\bibitem[{\citenamefont{Ward}(2005{\natexlab{b}})}]{ward2}
\bibinfo{author}{\bibfnamefont{B.~F.~L.} \bibnamefont{Ward}}
  (\bibinfo{year}{2005}{\natexlab{b}}), \eprint{hep-ph/0502104}.

\bibitem[{\citenamefont{Barrau et~al.}(2005)\citenamefont{Barrau, Feron, and
  Grain}}]{barrau05}
\bibinfo{author}{\bibfnamefont{A.}~\bibnamefont{Barrau}},
  \bibinfo{author}{\bibfnamefont{C.}~\bibnamefont{Feron}}, \bibnamefont{and}
  \bibinfo{author}{\bibfnamefont{J.}~\bibnamefont{Grain}},
  \bibinfo{journal}{Astrophys. J.} \textbf{\bibinfo{volume}{630}},
  \bibinfo{pages}{1015} (\bibinfo{year}{2005}).

\bibitem[{\citenamefont{Barrau et~al.}(2003)\citenamefont{Barrau, Blais,
  Boudoul, and Polarski}}]{barrau03}
\bibinfo{author}{\bibfnamefont{A.}~\bibnamefont{Barrau}},
  \bibinfo{author}{\bibfnamefont{D.}~\bibnamefont{Blais}},
  \bibinfo{author}{\bibfnamefont{G.}~\bibnamefont{Boudoul}}, \bibnamefont{and}
  \bibinfo{author}{\bibfnamefont{D.}~\bibnamefont{Polarski}},
  \bibinfo{journal}{Phys. Lett.} \textbf{\bibinfo{volume}{B551}},
  \bibinfo{pages}{218} (\bibinfo{year}{2003}).

\end{thebibliography}

\end{document}